\documentclass[aps,twocolumn]{revtex4}
\usepackage{float}

\usepackage{graphicx} 
\usepackage{amsmath,amssymb,amsfonts} 
\usepackage{url} 
\usepackage{multirow} 
\usepackage{float} 
\usepackage{lineno} 
\usepackage{xspace} 
\usepackage{ulem} 
\usepackage[usenames,dvipsnames]{color} 
\usepackage[colorlinks=true, linkcolor=blue, citecolor=blue, urlcolor=blue]{hyperref} 
\usepackage{float}

\newcommand{\bef}{\begin{figure}}
\newcommand{\eef}{\end{figure}}
\newcommand{\bc}{\begin{center}}
\newcommand{\ec}{\end{center}}

\newcommand{\be}{\begin{equation}}
\newcommand{\ee}{\end{equation}}
\newcommand{\bea}{\begin{eqnarray}}
\newcommand{\eea}{\end{eqnarray}}

\def\ba{\begin{eqnarray}}
\def\ea{\end{eqnarray}}

\begin{document}

\title{Collision Energy Dependence of Particle Ratios and Freeze-out Parameters in Ultra Relativistic Nucleus Nucleus Collisions}

\author{Iqbal Mohi Ud Din$^{*}$}
\affiliation{Department of Physics, Jamia Millia Islamia, New Delhi, India}

\author{Sameer Ahmad Mir}
\affiliation{Department of Physics, Jamia Millia Islamia, New Delhi, India}

\author{Nasir Ahmad Rather}
\affiliation{Department of Physics, Jamia Millia Islamia, New Delhi, India}

\author{Saeed Uddin$^{\dagger}$}
\affiliation{Department of Physics, Jamia Millia Islamia, New Delhi, India}

\author{Rameez Ahmad Parra}
\affiliation{Department of Physics, University of Kashmir, Srinagar, India}

\begin{abstract}
This work investigates the thermo-chemical freeze-out condition of the multi-component hot and dense hadron resonance gas (HRG) formed in the ultra-relativistic nucleus-nucleus collisions (URNNC). The van der Waals (VDW) type model used in the present analysis incorporates the  repulsive as well as attractive interactions among the hadrons. The baryons (antibaryons) are treated as incompressible objects. Using this theoretical approach the values of the model freeze-out parameters of the system are extracted over a wide range of collision energy by analyzing experimental data on like-mass antibaryon to baryon ratios.  The same set of parameters is found to explain the energy dependence of several other particle ratios quite satisfactorily.
We find that the horn-like structures seen in the ratios of strange particles to pions as a function of the collision energy cannot be explained by the VDW-HRG model alone without considering the strangeness imbalance effect in the system. We have compared our freeze-out line with those obtained earlier. The correlation between the  $\bar{p}/p$ and $K^-/K^+$ ratios is also examined.
\end{abstract}

\maketitle

\renewcommand{\thefootnote}{\fnsymbol{footnote}}

\footnotetext[2]{Corresponding Author E-mail: \href{mailto:suddin@jmi.ac.in}{suddin@jmi.ac.in}}
\footnotetext[1]{E-mail: \href{mailto:reshiiqbal24@gmail.com}{reshiiqbal24@gmail.com}}

\renewcommand{\thefootnote}{\arabic{footnote}} 

\section{INTRODUCTION}
\label{intro}
Understanding the quark-hadron phase transition diagram and the nuclear matter under extreme conditions requires investigating the thermodynamics of the hot and dense matter created in ultra-relativistic nucleus-nucleus collisions (URNNC). Exploring the behavior of matter a few microseconds after the Big Bang in the early universe and within the core of the neutron stars has gained significant attention.

The URNNC are a key tool for experimentally studying the high energy-density and high-temperature states of matter, such as the Quark-Gluon Plasma (QGP) in the laboratory. These collisions provide insights into collective effects, strange hadron yields and the possibility of a potential phase transition from a hadron resonance gas (HRG) phase, where quarks and gluons are confined, to a phase where they are deconfined or in a chirally symmetric phase. The URNNC programs at the Relativistic Heavy Ion Collider (RHIC) at BNL and the Large Hadron Collider (LHC) at CERN aim to study the properties of strongly interacting matter under extreme conditions. They also offer a unique opportunity to study the chemical equilibration of hadrons in the final state of collisions \cite{Muller1995,Adcox2005,Singh1992,Tannenbaum2006,Cleymans2011}.

 In URNNC the exploration of strongly interacting matter at extreme temperatures and densities is a focal point for both experimental and theoretical investigations.  Such studies have shown that statistical-thermal models can quite successfully reproduce the essential particle production features including their spectra in such collisions. The statistical hadronization model (SHM) reproduces transverse momentum spectra of different particle species, making it a useful tool for analyzing soft hadron production, particularly strangeness production which has been proposed as a signature of QGP formation \cite{Metag1993,Cleymans1986O,Letessier1993,Becattini1998,Munzinger1999,Yen1998,Parra2019S}.

Theoretical models, particularly the HRG model, have been instrumental in providing insights into the complex dynamics of strongly interacting particles. The Ideal Hadron Resonance Gas (IHRG) model, an initial formulation, treats mesons and baryons as point-like entities, assuming no interaction among them. However, deviations from an uncorrelated gas of hadrons are observed in lattice Quantum Chromodynamics (QCD) simulations \cite{Becattini2000,Munzinger2001,Florkowski2001,Becattini1995,Cleymans1999,Munzinger1994}. The ideal excluded volume type HRG models play a crucial role in understanding the quark-hadron phase transition and the particle production in URNNC \cite{Andronic2006}. In a Monte Carlo type approach with excluded-volume effect taken into account in the HRG, calculation of arbitrary moments of the event-by-event hadron yields have been done. However explaining the different types of particle ratios simultaneously over a wide range of energy is still lacking \cite{Bazavov2012}. It has also highlighted the systematic excess of experimental pion abundances when compared to the ideal hadron resonance gas results. This has also been attempted by assigning the pions a finite size \cite{Borsanyi2014}. These can lead to further review the various thermal and statistical descriptions of particle production in URNNC, including the excluded-volume effect in the HRG model.

Standard thermal model have generally no interactions among the hadrons after a stage called freeze-out. However, with no interaction taken into account during the evolution of the HRG till its final breakup stage, one cannot understand the hadron production in a realistic manner in URNNC. Hence, we need thermal models which incorporate interactions to study the properties of a hot and dense HRG including the relative hadronic yields in such a system \cite{Vovchenko2018M}. To address these challenges within the framework of a HRG model we can incorporate the attractive as well as the short-range repulsive interactions in van der Waals (VDW) type approach. Its ability to describe the nuclear liquid-gas transition at low temperatures and high baryon densities can set the stage for further useful exploration of the thermodynamic properties of HRG \cite{Yen1997,Tiwari2013,Mishra2008,Singh1991,Bashir2019P,Das2016C}.

For hot and dense hadron gas, several approaches have been proposed to formulate a precise equation of state (EOS), often using phenomenological models due to the lack of lattice QCD results at finite baryon density $(n_B)$. Thermal models, assuming thermo-chemical equilibrium, are effective in characterizing HRG, with the IHRG model being a basic version that treats hadrons and resonances as non-interacting point-like particles at freeze-out \cite{Welke1990}. It has been established that the short-range repulsion must be included to accurately describe HRG behaviour, as the IHRG model's phase transition calculations reveal inconsistencies at high temperatures and baryon densities \cite{Cleymans1986O,Cleymans1986Q}. The hadronic hard-core repulsive interactions become significant in a dense HRG and hence in a macroscopic approach this can be taken into account by assigning hadrons a geometrical hard-core volume, resulting in a van der Waals excluded-volume type effect that reduces particle densities in the system. Phenomenological models incorporating this effect \cite{Singh1991,Cleymans1987I,Kouno1989,Rischke1991,Anchishkin1992,Mir2023,Uddin2012,Parra2019,Mir2024,Bashir2018I,Parra2022M}  often lack a well-defined partition function or thermodynamic potential (\(\Omega\)), leading to thermodynamical inconsistencies in the calculations of particle densities, i.e.  \(n \neq \partial \Omega / \partial \mu\), where \(\mu\) is the chemical potential of the particle. It also breaks causality in dense HRG, making them less appealing despite their some empirical success.

Some earlier studies have also highlighted the importance of studying particle ratios as a function of the centre of mass energy in the URNNC. Hanafy  \cite{Hanafy2021} found that a statistically corrected HRG model can reasonably describe experimental results across different energies. Tawfik \cite{Tawfik2012P} also emphasized the significance of these ratios, particularly in the context of dynamical fluctuations in URNNC. Earlier Bjorken \cite{Bjorken1974} had also underscored the relevance of these ratios in understanding particle production in energetic collisions \cite{Blume2011,Tiwari2011,Tawfik2017,Stiles2006,Mir2024,Andronic2005,Bashir2015,Chatterjee2015,Mishra2008,Tiwari2013}.

The interesting horn-like structure in the kaon to pion ratio, as observed in URNNC, has been attributed to the release of a large number of colour degrees of freedom \cite{Nayak2005,Nayak2010,Nayak2011,Andronic2009Horn}. This phenomenon has been further explored in a hadronic kinetic model, which successfully reproduces the sharp peak in the \(K^+/\pi^+\) ratio \cite{Tomik2006}. These findings collectively suggest that the horn in the kaon to pion ratio is a result of the complex interplay of colour degrees of freedom and collision processes. However, in the present study we have attempted to investigate in detail this characteristic horn structure in the kaon-to-pion ratio as well as other strange hadron to pion ratios in URNNC. Within the framework of the model used in the present study we have found strong evidence of strangeness imbalance factor, $\xi_s$.  This helps us in understanding the production of strange particles relative to pions in the system 
produced in URNNC \cite{Bir1982,Grebieszkow2009,Bhattacharyya2017,Tawfik2021H,Cleymans2017T,Oeschler2017,Cleymans2004T}.

The present work is organized as follows. In Section~\ref{sec:formulation} and ~\ref{sec:vdw} we have provided some basic idea of the model formulation. In Section~\ref{sec:Results} the results of our calculations and their comparison with the experimental data are presented along with the discussion. Finally in Section~\ref{Sec:sum} we have summarized our results.

\section{STATISTICAL THERMAL MODEL}
\label{sec:formulation}
A simple and popular model for describing the pressure function for a system of particles in equilibrium having both repulsive and attractive interactions is the one described by the van der Waals (VDW) equation of state. In canonical ensemble representation, it can be written as \cite{Samanta2017C,Greiner1965,Landau1980}:

\begin{equation}
\left[P+\left(\frac{N}{V}\right)^2a\right]\left(V-b\right)=NT
\end{equation}
where the attractive and repulsive interactions between the particles are described, respectively, by the positive VDW parameters \(a\) and \(b\). The pressure, volume, temperature, and the number of particles having an “effective finite-size” due to their hard-core repulsion among them in the system are represented by the letters \(P\), \(V\), \(T\), and \(N\), respectively. The above can be rewritten in a simple and transparent form as:

\begin{equation}
p\left(T,n\right)=\frac{NT}{V-bN}-a\frac{N^2}{V^2}=\frac{nT}{1-bn}-an^2
\label{eq:2}
\end{equation}
The symbol \(n\) represents the finite-size particle number density. The first term includes the correction factor due to hard-core repulsion by assigning a hard-core radius to the particles (baryons and antibaryons in our case). Considering particles as rigid non-deformable spheres, we have \(b=16\pi r^3/3\), where \(r\) is the particle's hard-core radius. The second term accounts for the attractive interactions between particles through the parameter \(a\). For \(a=0\), the equation reduces to the EV-HRG equation, containing only repulsive interactions, while for both \(a=0\) and \(b=0\) it reduces to the ideal HRG case \cite{Greiner1965,Tiwari2011,Bugaev2000,Sahoo2023,Yen1997}.

In a statistical approach, it is necessary to adopt the grand canonical ensemble (GCE) formulation to study the systems formed in the URNNC \cite{Samanta2017C,Vovchenko2018Q,Vovchenko2017M,Pradhan2022,Vovchenko2015P,Sarkar2018,Vovchenko2015S,Vovchenko2015V}. The preference for the Grand Canonical Ensemble (GCE) over the Canonical Ensemble (CE) stems from its technical simplicity in describing the properties of the systems due to its non-conserved (fluctuating) particle numbers in the system which is essentially due to the particle production/annihilation reactions taking place inside it. The GCE approach is used to calculate the mean particle densities in the HRG system. The CE formulation on the other hand depends on parameters like volume (\(V\)), temperature (\(T\)), energy (\(E\)) and the particle number (\(N\)) of the system. In the GCE formulation, we can calculate mean particle number density (\(n\)), energy density (\(\varepsilon\)), entropy density (\(s\)), etc., which are essential for understanding the system’s behaviour. Here we will apply the following statistical approach, which was earlier used for the EV model, i.e., for \(a=0\) in Eq. (2), \cite{Rischke1991,Gorenstein1981,Parra2019,Parra2023,Parra2019T,Din2024}.

\section{ VDW EQUATION OF STATE THROUGH GRAND CANONICAL ENSEMBLE:}
\label{sec:vdw}

We discuss the grand canonical ensemble (GCE) formulation by considering both attractive and repulsive interactions. The repulsive force is assumed to exist between pairs of two baryons and pairs of two antibaryons, while it is purely attractive between a baryon-antibaryon pair as the annihilation processes predominantly govern short-range attractive interactions between baryon-antibaryon pairs. The hard-core repulsive interaction among bosons (mesons) has been neglected \cite{Vovchenko2016V,Vovchenko2015V,Samanta2017C,Andronic2012}.

The grand canonical partition function (\(Z_{GC}\)) for a system can be written as:

\begin{equation}
Z_{GC}\left(T,\mu,V\right)=\sum_{N=0}^{\infty} e^{\mu N/T} Z_C\left(T,N,V\right)
\end{equation}

where \(Z_C\left(T,N,V\right)\) stands for the canonical partition function applicable to a system with \(N\) number of particles. \(T\) signifies the temperature and \(V\) represents the total physical volume of the system. The phenomenological grand partition function, incorporating both attractive and repulsive interactions, can be formulated as follows:

\begin{equation}
Z_{GC}^{int}\left(T,\mu,V\right)=\sum_{N=0}^{\infty} e^{\mu N/T} Z_C\left(T,N,V-bN\right) e^{-\bar{U}/T}
\label{eq:4}
\end{equation}

Here, \(Z_C\left(T,N,V-bN\right)\) is the canonical partition function accounting for the excluded volume effect due to hard-core repulsion. The term \(bN\) represents the total excluded volume resulting from the hard-core repulsive interaction among baryons (or among antibaryons) \cite{Rischke1991,Uddin1994,Kouno1989}. The attractive interaction is incorporated by introducing the factor \(e^{-\bar{U}/T}\) in the grand partition function, where \(\bar{U}\) represents the average attractive interaction energy which can be expressed as

\begin{equation}
\bar{U} = \frac{1}{2} \sum_{i,j} V_{att}\left(\vec{r_i} - \vec{r_j}\right)
\end{equation}

The summation over the indices \(i, j\) runs over all particles (\(N\)) in the system. The \(V_{att}\left(\vec{r_i} - \vec{r_j}\right)\) represents the mean attractive interaction energy between any two particle pairs. Interactions involving three or more particles are disregarded due to their rarity. Assuming a uniform finite-size particle number density, \(n = \frac{N}{V}\), the total interaction energy can be determined by summing over all particle pairs. Therefore, for large \(n\), we can express it in the integral form as follows\cite{Landau1980,Armiento2005,Ramshaw1980}:

\begin{equation}
\bar{U} = \frac{1}{2} \int d^3\vec{r_1} d^3\vec{r_2} \, n\left(\vec{r_1}\right) n\left(\vec{r_2}\right) V_{att}\left(\vec{r_2} - \vec{r_1}\right)
\end{equation}

\noindent Finally, this gives \(\bar{U} \approx n^2 V a\), where \(a = \int 2\pi r^2 V_{att}\left(\vec{r}\right) d\vec{r}\) and can be regarded as an effective attractive interaction parameter. In view of this, we can rewrite Eq.~\ref{eq:4} as:
\begin{equation}
Z_{GC}^{int}\left(T,\mu,V\right)=\sum_{N=0}^{\infty} e^{\mu N/T} Z_C\left(T,N,V-bN\right) e^{-n^2Va/T}
\end{equation}

Defining the grand partition function involving only the hard-core repulsive interaction as:

\begin{equation}
Z_{GC}^{excl}\left(T,\mu,V\right)=\sum_{N=0}^{\infty} e^{\mu N/T} Z_C\left(T,N,V-bN\right)
\end{equation}

we can write:

\begin{equation}
Z_{GC}^{int}\left(T,\mu,V\right) = Z_{GC}^{excl}\left(T,\mu,V\right) e^{-\frac{n^2Va}{T}}
\label{eq:9}
\end{equation}

In order to define the system’s pressure, we adopt a standard procedure wherein we take the Laplace transformation of \(Z_{GC}^{int}\left(T,\mu,V\right)\) \cite{Rischke1991,Parra2019,Parra2023}:

\begin{equation}
\hat{Z}_{GC}^{int}\left(T,\mu,\zeta\right) = \int e^{-\zeta V} Z_{GC}^{int}\left(T,\mu,V\right) dV
\end{equation}

In order to overcome the extreme right-hand singularity, the above integral must be convergent in the limit \(V\rightarrow\infty\). This provides:

\begin{equation}
\zeta = \lim_{V\rightarrow\infty} \left[\frac{\ln Z_{GC}^{int}\left(T,\mu,V\right)}{V}\right]
\end{equation}

or

\begin{equation}
\zeta = \frac{1}{T} \lim_{V\rightarrow\infty} \left[\frac{T \ln Z_{GC}^{int}\left(T,\mu,V\right)}{V}\right]
\end{equation}

This gives:

\begin{equation}
\zeta = \frac{1}{T} p^{int}\left(T,\mu,V\right)
\end{equation}
We can again write the Laplace transform using Eq.~\ref{eq:9} as:

\begin{equation}
\hat{Z}_{GC}^{int}\left(T,\mu,\zeta\right) = \int e^{-\zeta V} Z_{GC}^{excl}\left(T,\mu,V\right) e^{-\frac{n^2Va}{T}} dV
\label{eq:15}
\end{equation}

Rewriting:

\begin{equation}
\hat{Z}_{GC}^{int}\left(T,\mu,\zeta\right) = \int e^{-V\left(\zeta - \frac{\ln Z_{GC}^{excl}\left(T,\mu,V\right)}{V} + \frac{n^2a}{T}\right)} dV
\end{equation}
and
\begin{equation}
\zeta = \frac{1}{T} \lim_{V\rightarrow\infty} \left[\frac{T \ln Z_{GC}^{excl}\left(T,\mu,V\right)}{V} - n^2a\right]
\end{equation}
This leads to:
\begin{equation}
p^{int}\left(T,\mu,V\right) = p^{excl}\left(T,\mu,V\right) - an^2
\label{eq:17}
\end{equation}

When a=0, the $p^{excl}\left(T,\mu,V\right)$ for the case of GCE turns out to be:  \cite{Gorenstein1999V,Rischke1991,Zeeb2002E}

\begin{equation}
p^{excl}\left(T,\mu,V\right) = \frac{nT}{1-bn}
\label{eq:18}
\end{equation}

where \(n\) in this case is the GCE averaged particle number density. This gives for Eq.~\ref{eq:17} the following:

\begin{equation}
p^{int}\left(T,\mu,V\right) = \frac{nT}{1-bn} - an^2
\label{eq:19}
\end{equation}

In principle, the particle number density can be obtained in a thermodynamically consistent manner as \(n = \left.\frac{\partial p^{int}}{\partial \mu}\right|_T\) \cite{Parra2019,Parra2023}:

\begin{equation}
n = \left(\frac{T}{\left(1-bn\right)^2} - 2an\right) \left.\frac{\partial n}{\partial \mu}\right|_T
\end{equation}

After further calculations, one can obtain an expression for the chemical potential, \(\mu\), as:

\begin{equation}
\mu = T \ln\left(\frac{n}{1-bn}\right) + \frac{T}{1-bn} - 2an + C
\end{equation}

For \(a=0, b=0\) we should recover the ideal case, i.e., \(n(T,\mu)=n_{\mathrm{id}}(T,\mu)\) where we can write (using Boltzmann approximation) \(n_{\mathrm{id}}\left(T,\mu\right) = \phi\left(T,m\right) \exp\left(\frac{\mu}{T}\right)\) for the point-like particle number density. The integration constant \(C\) comes out to be \cite{Gorenstein2022,Vovchenko2015P,Parra2023,Kostyuk2000}.

\begin{equation}
C = T \ln\left(\frac{n_{\mathrm{id}}}{\phi}\right) - T \ln n_{\mathrm{id}} - T
\end{equation}

Thus we can write:

\begin{equation}
\mu = T \ln\left(\frac{n}{(1-bn)\phi}\right) + \frac{Tnb}{1-bn} - 2an
\end{equation}

Defining:

\begin{equation}
\mu^\ast = T \ln\left[\frac{n}{(1-bn)\phi}\right]
\end{equation}

and again writing the quantity \(\phi\) in terms of \(\mu^\ast\) as:

\begin{equation}
\phi = n^{\mathrm{id}}\left(T,\mu^\ast\right) e^{-\frac{\mu^\ast}{T}}
\label{eq:25}
\end{equation}
we get:
\sloppy
\begin{equation}
\mu^\ast = \mu - \frac{bnT}{1-bn} + 2an
\end{equation}
Using Eq.~\ref{eq:15} we can finally write the above result as:

\begin{equation}
\mu^\ast = \mu - bp^{excl}\left(T,\mu,V\right) + 2an
\end{equation}

Combining Eq.~\ref{eq:18} and Eq.~\ref{eq:19} we will get:

\begin{equation}
n(T,\mu) = (1-bn) n^{\mathrm{id}}\left(T,\mu^\ast\right)
\end{equation}

which finally yields for the number density of the particle \(n\left(\mu,T\right) = n^{int}(T,\mu)\) as:

\begin{equation}
n\left(\mu,T\right) = \frac{n^{\mathrm{id}}\left(T,\mu^\ast\right)}{1+bn^{\mathrm{id}}\left(T,\mu^\ast\right)}
\end{equation}

In view of the result in Eq.~\ref{eq:19}, the above Eq.~\ref{eq:25} takes the form of a transcendental equation.
For a multi-component system, it is possible to generalize the above result for a given \(j\)th particle species as \cite{Zeeb2002E}:

\begin{equation}
n_j\left(\mu_j,T\right)=\frac{{n^{\mathrm{id}}}_j\left(T,\mu_j^\ast\right)}{1+\sum_{i}{b{n^{\mathrm{id}}}_i\left(T,\mu_j^\ast\right)}}
\end{equation}
\fussy

The summation over the index \(i\) is for all species having hard-core repulsive interaction, including the \(j^{th}\) specie. Further details can be found in \cite{Parra2022M,Vovchenko2016V,Vovchenko2015P}. The above results can be effectively utilized to calculate relative particle yields in the system.

The resonances that decay into specie \(j\) also contribute to its finally measured yield in the actual experiments. 
Therefore, the contributions from all heavier hadrons (say \(k^{th}\) up to the mass 2 GeV) that decay into hadron \(j\) with their known branching fraction \(\Gamma_{k \rightarrow j}\) have been taken into account \cite{ParticleDataGroup2020} to give us:


\begin{equation}
n_j\left(\mu_j,T\right)={n_j\left(\mu_j,T\right)}^{prim}+\sum_{k\neq j}{\Gamma_{k\rightarrow j}n_k\left(\mu_k,T\right)}
\end{equation}

Where \(\Gamma_{k \rightarrow j}\) is the branching ratio for the decay of the $k^{th}$ resonance into the $j^{th}$ particle state. The superscript \(prim\) stands for the primary number density without decay contribution. The summation is over all the heavier resonances \(k\) that decay to the $j^{th}$ hadron. Here we have considered the decay contributions to the final state hadron multiplicities from single weak decays of the heavy hadronic resonances as well as those double decays where a weak decay is followed by a strong decay.

In the present analysis we have treated \textit{b} as a free parameter which is fitted to observed particle ratios, yielding values in close agreement with
estimates based on lattice QCD data. \cite{Tiwari2011,Samanta2017C, Vovchenko2017E,Albright2014M,Mishra2007Effect,Sarkar2018}. The value of the parameter \(a = 329 \, \text{MeV} \ \text{fm}^3\) of the model is chosen from previous works  \cite{Vovchenko2015V,Bethe1971,Cleymans1986Q,Wang2022,Greiner1965,Bugaev2000,Zhang1992}

\section{RESULTS AND DISCUSSION}
\label{sec:Results}
In this section, we present the outcome of our analysis of particle production, including the strangeness imbalance effect observed in URNNC over a wide range of collision energy, using the above discussed VDW-HRG model. 

For the \(i^{\text{th}}\) hadron having baryon number \(B_i\), strangeness \(S_i\), and electric charge \(Q_i\), we associate baryon chemical potential \(\mu_B\), strange chemical potential \(\mu_S\), and electric chemical potential \(\mu_Q\), respectively. These are treated as model parameters which control the net baryon, strangeness, and electric charge content of the system. The overall chemical potential of the \(i^{\text{th}}\) hadron is thus defined as:
\begin{equation}
\mu_i = B_i\mu_B + S_i\mu_S + Q_i\mu_Q
\end{equation}

The strangeness conservation criteria and the initial charge-to-baryon ratio of the colliding system can be used to fix \(\mu_S\) and \(\mu_Q\), respectively, as these quantities remain conserved during the formation and subsequent evolution of the hot hadronic system created in URNNC \cite{Biswas2020}. We can thus write the following \cite{Chatterjee2015}:

\begin{equation}
\sum_{i} n_i\left(T, \mu_B, \mu_S, \mu_Q\right) S_i = \text{Net } S = 0
\end{equation}
\label{eq:32}
\begin{equation}
\sum_{i} n_i\left(T, \mu_B, \mu_S, \mu_Q\right) Q_i = \mathcal{R} \sum_{i} n_i\left(T, \mu_B, \mu_S, \mu_Q\right) B_i
\end{equation}

Here, \(\mathcal{R} = \text{Net } Q / \text{Net } B\) is the net charge-to-net baryon number ratio of the colliding nuclei. For nucleus-nucleus collisions like Au+Au, Pb+Pb, Ag+Ag, etc., we find \(\mathcal{R} = N_p / (N_p + N_n) \approx 0.4\), with \(N_p\) and \(N_n\) denoting the number of protons and neutrons in the colliding nuclei. The  \text{Net} \textit{S} is total strangeness of the system.

The freeze-out parameters of the system formed in URNNC, such as temperature and baryon chemical potential, are influenced by various factors, especially the collision energy \cite{Tiwari2013} hence, to describe the variations in particle ratios across a wide range of energies up to the LHC, we need to understand how the baryon chemical potential and temperature at the thermo-chemical freeze-out may depend on the collision energies. We have adopted a standard methodology used to fix the temperature \(T\) and the baryon chemical potential \(\mu_B\) of the system for different center-of-mass energies, \(\sqrt{s_{NN}}\). Numerous studies \cite{Bashir2015,Cleymans2006C,Mishra2008,Tiwari2013,Andronic2006,Cleymans1999,Chatterjee2015,Biswas2020,Tawfik2012P} have suggested a suitable ansatz-like dependence of \(T\) and \(\mu_B\) on \(\sqrt{s_{NN}}\). Therefore, following the same pattern, we employ the following parameterization \cite{Cleymans2006C}:

\begin{equation}
\mu_B = \frac{c}{1 + d\sqrt{s_{NN}}}
\label{eq:35}
\end{equation}

\begin{equation}
T = e - f\mu_B^2 - g\mu_B^4
\label{eq:36}
\end{equation}

To ascertain the values of the parameters (\(c, d, e, f, g\)), we use the best fit criteria, i.e., by obtaining the minimum value of the chi-squared per degree of freedom (\(\chi^2/\text{\textit{dof}}\)), to the experimentally measured antibaryon to baryon ratios over a wide range of collision energy, \(\sqrt{s_{NN}}\). The chi-square (\(\chi^2/\text{\textit{dof}}\)) is computed as:

\begin{equation}
\frac{\chi^2}{\text{\textit{dof}}} = \frac{1}{N}\sum_{i=1}^{N}\frac{\left(R_i^{\text{the}} - R_i^{\text{exp}}\right)^2}{\sigma_i^2}
\end{equation}

where \( R_i^{\text{the}} \) and \( R_i^{\text{exp}} \) represent theoretical and experimental values, respectively, with \( \sigma_i \) as the experimental uncertainties in the measured quantities at corresponding \( \sqrt{s_{NN}} \), with \( N \) as the number of degrees of freedom in the data set. The values of the model parameters are extracted from the best fit method and are found to be  :
\vspace{-10mm}
\begin{center}
\begin{align*}
& c = 1398 \pm 29\, \text{MeV}, \quad d = 0.28 \pm 0.008\, \text{GeV}^{-1}, \\
& e = 164 \pm 2.5\, \text{MeV}, \quad f = 0.14 \pm 0.021\, \text{MeV}^{-1}, \\
& g = 0.015 \pm 0.04\, \text{MeV}^{-3}
\end{align*}
\end{center}

The values of the above parameters along with that of the baryonic hard-core radius, \textit{r} = \( 0.73 \pm 0.02 \, \text{fm} \), provide (nearly) simultaneous minimum \( \chi^2/\text{\textit{dof}} \) fit to all four antibaryon to baryon ratios (i.e., \( \bar{p}/p \), \( \bar{\Lambda}/\Lambda \), \( \bar{\Xi}^-/\Xi^- \) and \( \bar{\Omega}/\Omega \)). This estimate of the baryonic hard-core radius is in close agreement with the earlier estimate based on the description of the lattice QCD data of different thermodynamics quantities \cite{Samanta2017C}.

\subsection{LIKE MASS ANTIBARYON TO BARYON RATIOS}
\begin{figure} 
    \centering
    \includegraphics[width=0.45\textwidth]{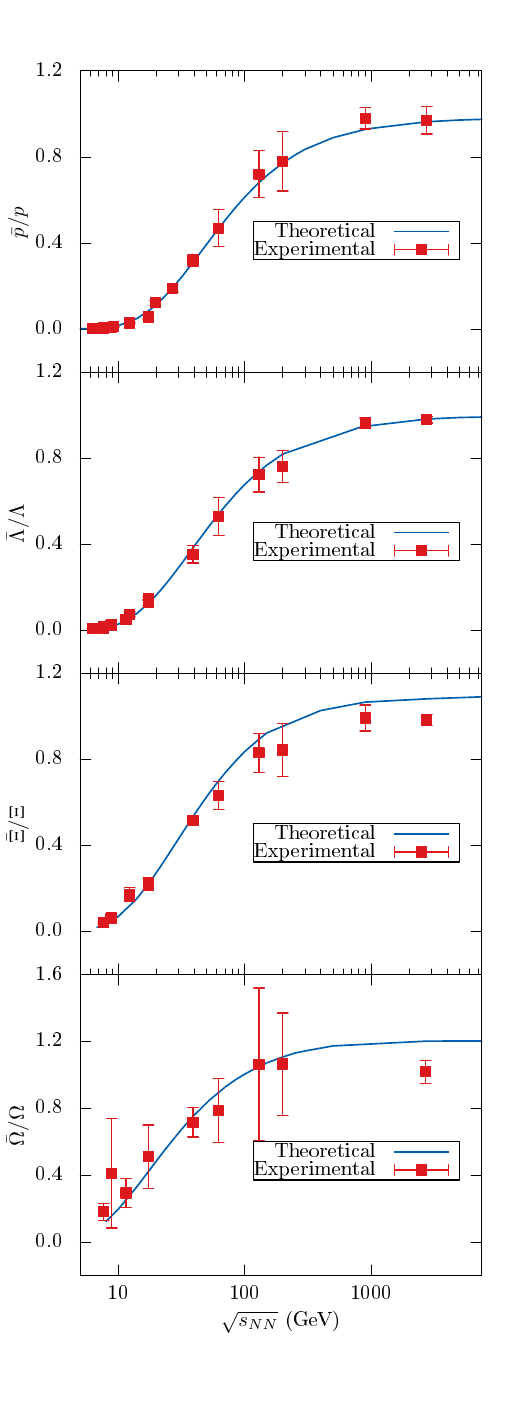} 
    \caption{Collision energy dependence of \( \bar{p}/p \), $\bar{\Lambda}/\Lambda$, $\bar{\Xi}/\Xi$, $\bar{\Omega}/\Omega$.}
    
    \label{fig:1}
\end{figure}

Experimental data show that all antibaryon-to-baryon ratios increase with \(\sqrt{s_{NN}}\) and approach unity as we move towards the LHC energies, where the temperature \((T)\) almost saturates at 164 MeV and the chemical potential \((\mu_B\)) nearly vanishes, indicating an almost baryon-antibaryon symmetric matter. At lower energies (i.e., SPS and AGS), the system maintains a large excess of baryons over antibaryons, resulting in all the antibaryon-to-baryon ratios being significantly lower than unity \cite{Grebieszkow2004}.
In Fig.~\ref{fig:1}, we have shown the energy dependence of antibaryon to baryon ratios, i.e., \( \bar{p}/p \), \( \bar{\Lambda}/\Lambda \), \( \bar{\Xi}/\Xi \), and \( \bar{\Omega}/\Omega \). The minimum $\chi^2/\text{\textit{dof}}$ values for the above four cases are 0.54, 0.79, 3.92, and 0.68, respectively. A relatively larger value in case of \( \bar{\Xi}/\Xi \) is due to the small error bar associated with the data point at 2.76 TeV.

The increasing antibaryon to baryon ratios with $\sqrt{s_{NN}}$ also suggests a higher degree of nuclear transparency in the URNNC at higher collision energies, where the bulk of the secondary matter is formed between the two receding nuclei due to high excitation of vacuum and hence is almost baryon symmetric. This effect becomes more prominent as we move towards the highest RHIC and further up to the LHC energies. In lower energy collisions, the bulk secondary matter contains a large contribution from the participant nucleons due to significant baryon stopping in these collisions. This observation is also supported by the decreasing value of net-protons ($p - \bar{p}$) measured in the experiments with increasing collision energy. The matter formed at lower RHIC energies indicates a situation where partial transparency exists in the nuclear collisions. This observation can provide insights into the evolving collision dynamics with increasing energy \cite{Abelev2010}.
The curves obtained by our VDW-HRG model calculations exhibit good agreement with experimental data taken from various available sources 
\cite{Adamczyk2017B,Ahle1999P,Adams2002S,Aggarwal2011S,Aamodt2010M, Afanasiev2002,Antinori2004,Appelshauser1998,Alt2005Omega,Mischke2002,Abbas2013b,Abelev2010}. The energy-dependent behavior of antihyperon/hyperon ratios, depicted in the  Fig.~\ref{fig:1}
reveals a mass-hierarchical trend, with a saturation value of 1 being reached earlier for more massive hyperon species. This phenomenon underscores a preference for almost symmetric antihyperons-hyperons production in the case of multistrange hyperons, where the former, i.e., the $\Lambda$, contains only two light valence quarks, while the $\Xi$ contains only one light valence quark, the likely remnants from the initial nucleons. In the case of $\mathrm{\Omega}$, all (strange) valence quarks are newly generated \cite{Andronic2006, Bashir2015}. The generally unsatisfactory quality of thermal fits to the multi-strange baryons data at SPS energies cannot solely be attributed to these cases, as we shall see below \cite{Andronic2006}. All these ratios increase sharply with respect to $\sqrt{s_{NN}}$ and then almost saturate at higher energies (above 200 GeV), reaching a value equal to 1.0 at LHC energy. This behavior shows that the production rates of anti-particles relative to particles continuously increase with increasing $\sqrt{s_{NN}}$ and become almost equal at LHC energy. The excellent agreement between our VDW-HRG model results and the experimental data validates the ability of the model to explain these particle ratios.
\subsection{OTHER PARTICLE RATIOS}

\begin{figure}[htbp]
    \centering
    \includegraphics[width=1.0\columnwidth]{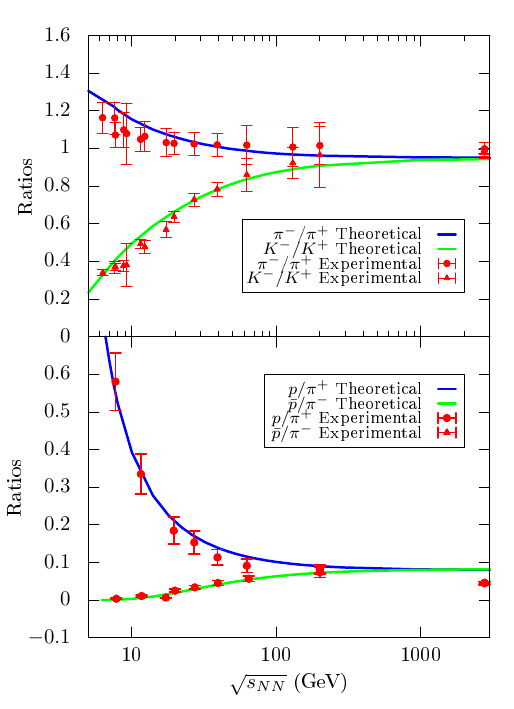} 
    \caption{$\sqrt{s_{NN}}$ dependence of $\pi^{-}/\pi^{+}$, $K^{-}/K^{+}$} and \( p/\pi \)
    \label{fig:2}

\end{figure}

We extend our analysis to further test the above values of the chemical freeze-out parameters. We find that the $K^-/K^+$ ratio is well described with the same values of the model parameters used to obtain the best fitted curves for antibaryon to baryon ratios. In Fig.~\ref{fig:2} the $K^-/K^+$ ratio data against $\sqrt{s_{NN}}$ is well fitted by the curve obtained by using the VDW-HRG model calculations with \( \chi^2/\text{\textit{dof}} \) = 1.07 \cite{Abelev2010,Abelev2009,Acharya2020,Adams2002S,Abelev2013,Mischke2002}. The decline in the $K^-/K^+$ ratio as collision energy decreases is attributed to a rise in net baryon density, favoring the production of $K^+$ (having $u$ and $\bar{s}$ valence quarks) over $K^-$ (having $\bar{u}$ and $s$ valence quarks). As the collision energy increases, the $K^-/K^+$ ratio exhibits a systematic rise and approaches unity, indicative of the pair production processes for $K^+$ and $K^-$ at higher energies in almost equal proportion through the reaction channels involving baryons and antibaryons in an almost baryon symmetric matter (having $\mu_{B} \sim 0$). The associated production mechanism yields $K^+$ through reactions like $\pi + N \rightarrow Y + K^+$ and the production of $K^-$ takes place through $\pi + \bar{N} \rightarrow \bar{Y} + K^-$, where $N$ represents a nucleon and $Y$ denotes a singly strange hyperon. On the other hand, reactions like $N + \bar{N} \rightarrow K^- + K^+$ and $\pi^+ + \pi^- \rightarrow K^- + K^+$ will produce $K^+$ and $K^-$ in equal proportion \cite{Das2014}.
For the purpose of comparison and using the same set of parameters we have also shown the energy dependence of the $\pi^-/\pi^+$ ratio in the same figure where both ratios approach $\sim 1$ for large $\sqrt{s_{NN}}$, as the system tends to become (almost) electric charge and baryon symmetric due to nuclear transparency effect. The green curve shows the theoretical fit, which matches the experimental data \cite{Bratkovskaya2004,Adams2002S,Abelev2013,Adamczyk2017B,Adler2002} quite well with $\chi^2/\text{\textit{dof}} = 1.25$. It may be recalled here that in our calculation we have  applied the criteria of conserved charge to baryon ratio in the experiments in order to fix the electric chemical potential $\mu_Q$. We find that at low beam energies, the $\pi^-/\pi^+$ ratios exhibit values larger than unity, possibly due to significant contributions to the positive charges from protons and other resonances, particularly the $\Delta$ baryons. 
Further, since protons (antiprotons) are the lightest baryons (antibaryons) hence most high-mass baryons (antibaryons) decay into protons (antiprotons). The $p/\pi$ ratio, therefore, can characterizes well the overall baryon (antibaryon) production relative to the pions which are copiously produced in URNNC. To understand this, in Fig.~\ref{fig:2} we have shown the $p/\pi^+$ and $\bar{p}/\pi^-$ ratios energy dependence. The theoretical curve obtained for the same set of the parameters values appears to fit the data quite well with $\chi^2/\text{\textit{dof}} = 0.71$ and $3.09$, respectively \cite{Nandi2020,Aggarwal2011S,Adamczyk2017B,Abelev2010}. The relatively larger value of $\chi^2/\text{\textit{dof}}$  for  the $\bar{p}/\pi^-$ case results due to very small error bars in the experimental data.

\begin{figure}[H]
    \centering
    \includegraphics[width=0.98\columnwidth]{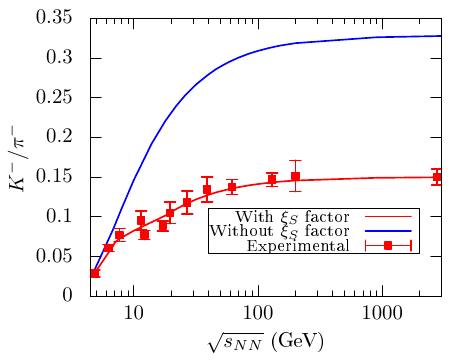} 
    \caption{Variation of $K^{-}/\pi^{-}$ with $\sqrt{s_{NN}}$}
    \label{fig:3}
\end{figure}

\begin{figure}[htbp]
    \centering
    \includegraphics[width=0.98\columnwidth]{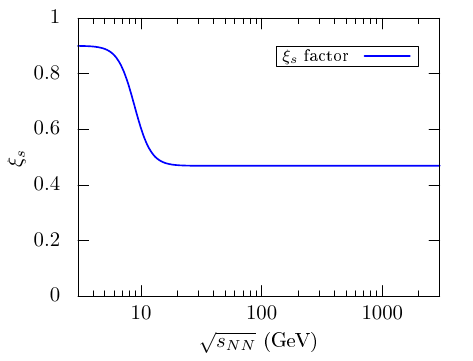} 
    \caption{\centering
        \parbox{0.8\linewidth}{\centering 
        $\sqrt{s_{NN}}$ dependence of the strangeness imbalance factor $\xi_s$}
    }
    \label{fig:4}
\end{figure}

Despite the success of the HRG Model in describing many particle ratios, some unresolved issues still remain while describing the yields of the strange (antistrange) particles relative to the non-strange ones. For nearly a decade, the $K/\pi$, $\Lambda/\pi^-$, $\Xi/\pi^-$, $\Omega/\pi^{-}$, $\Phi/\pi^-$ and $\Phi/K^{-}$ ratios were not accurately described \cite{Beutler2008,Tawfik2016}. 
Pions being the lightest hadrons are produced copiously and carry the bulk of the entropy of the system.
The dependence of the $K/\pi$ on $\sqrt{s_{NN}}$ has always been a subject of interest as it is a crucial indicator of the strangeness content of the system relative to its entropy in heavy-ion collisions and has implications for the quark-hadron phase transition. The observed increase and subsequent saturation of the $K/\pi$ ratio with increasing beam energies align with established trends and theoretical expectations \cite{Andronic2006}. As already mentioned we have assumed simultaneous freeze-out for all particles in our calculation. The resulting fit to the experimental $K^-/\pi^-$ ratio's dependence on $\sqrt{s_{NN}}$ is shown in Fig.~\ref{fig:3} in blue color. It is obvious that the theoretical curve provides an unsatisfactory fit to the experimental data \cite{Adams2002S,Abelev2010,Abelev2009,Abelev2013,Adamczyk2017B,Adler2002}. However, such large disagreements have been also reported earlier which is attributed to strangeness imbalance in the system \cite{Castorina2016,Munzinger2003,Becattini2005,Bugaev2016}.


\begin{figure}[htbp]
    \centering
    \includegraphics[width=0.98\columnwidth]{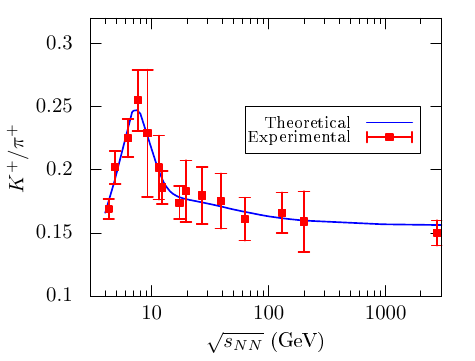} 
    \caption{Variation of $K^+/\pi^+$ with $\sqrt{s_{NN}}$}
    \label{fig:5}
\end{figure}

The theoretical curve is seen to highly over-predict the experimental data at higher collision energies ($\sqrt{s_{NN}} > 10 \, \text{GeV}$). To explain this disagreement and further analyze other strange to non-strange particle ratios, we introduce a strangeness imbalance factor $\xi_s$ which is defined at different energies (say for the $i^{th}$) as follows \cite{Tawfik2021H,Biswas2020}:
\begin{equation}
\xi_s^i = \frac{(K^- / \pi^-)_{\text{exp}}^i}{(K^- / \pi^-)_{\text{theo}}^i}
\end{equation}

In order to make it convenient to apply this factor to other particle ratios as well for any given collision energy, we have used a parameterization method to represent the energy dependence of $\xi_s$. We have used the following ansatz to best represent the obtained $\sqrt{s_{NN}}$ dependent values of $\xi_s$:
\begin{equation}
\xi_s = \alpha + \beta \left(1 + \left[\frac{\sqrt{s_{NN}}}{\gamma}\right]^\delta\right)^{-1}
\label{eq:39}
\end{equation}

The best spline fit provides: 
\vspace{-5mm}
\begin{center}
    
$\alpha = 0.469\pm0.016$, $\beta = 0.432\pm0.006$, \\
$\gamma = 8.83\pm0.364 \, \text{GeV}$, and $\delta = 6.71\pm0.36$.
\end{center}

\vspace{-5mm}

In Fig.~\ref{fig:4}, we have shown the energy dependence of $\xi_s$. Utilizing the ansatz of Eq.~\ref{eq:39}, we have redefined the strangeness imbalance corrected strange (antistrange) particle number densities in our calculation at different collision energies as
\begin{equation}
n_s^i\left(\text{corrected}\right) = \left[\xi_s^i\right]^h n_s^i\left(\text{uncorrected}\right)
\end{equation}

The quantity $h$ represents the number of strange and/or antistrange quarks in the given hadron. For singly strange particles (i.e., kaons, antikaons, lambda, antilambda, sigma, antisigma) we have $h = 1$; for doubly strange particles (i.e., cascades, anticascades) as well as $\phi$ meson we take $h = 2$; and similarly for triply strange particles (i.e., omega, antiomega) we have $h = 3$. In Fig.~\ref{fig:3}, the corrected theoretical curve for the $K^-/\pi^-$ ratio using the above ansatz in Eq.~\ref{eq:39} is shown by the red curve, which has a very small $\chi^2/\text{\textit{dof}} = 0.51$, resulting from the correction procedure. Moreover, using the values of $\xi_s$ obtained from Eq.~\ref{eq:39}, we have also calculated the strangeness imbalance corrected $K^+/\pi^+$ ratio for different collision energies. In Fig.~\ref{fig:5}, the theoretical curve for the $K^+/\pi^+$ ratio is shown in blue, having value of $\chi^2/\text{\textit{dof}} = 0.15$. Interestingly, the theoretical curve also explains the horn structure at lower energies, $\sqrt{s_{NN}} \sim 7-8 \, \text{GeV}$. The rapid variation of the $(K^+/\pi^+)$ and $(K^-/\pi^-)$ ratios at lower energies reflects the threshold for strangeness production affected by the rapid initial increase in the temperature with $\sqrt{s_{NN}}$. A steep drop in the $K^+/\pi^+$ near the horn may arise due to the rapidly falling baryon chemical potential, as baryon-rich matter favors $K^+$ production, as discussed above. Though the rapid rise in temperature and the subsequent drop in baryon chemical potential with increasing collision energy are significant factors, they alone cannot fully explain the horn structure within the VDW-HRG model without the presence of the strangeness imbalance factor $\xi_s$ which is critical in capturing the observed behavior. However, due to the strangeness imbalance effect the like-mass anti-hyperon to hyperon as well as antikaon to kaon ratios are not modified, as their abundances in the system are equally affected and in their theoretical calculations the correction factor $\xi_s$ cancels out.


\begin{figure}[H]
    \centering
    \includegraphics[width=0.98\columnwidth]{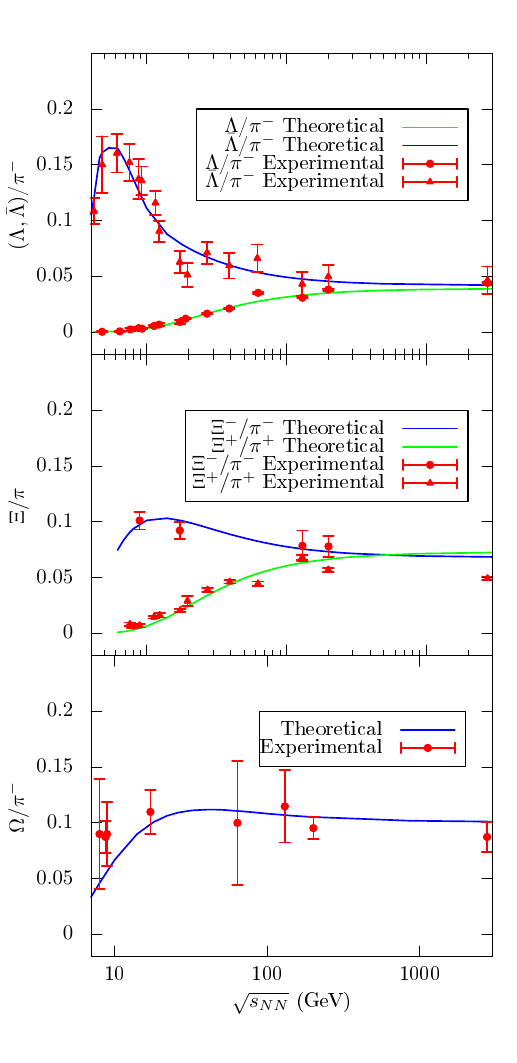} 

    \caption{Dependence of  $\Lambda (\bar{\Lambda})/\pi^-$, $\Xi/\pi$, and $\Omega/\pi^{-}$ on $\sqrt{s_{NN}}$}
    .
   \label{Fig:6}

\end{figure}


The observation that the $K^-/\pi^-$ and $K^+/\pi^+$ ratios in Figs.~\ref{fig:3} and~\ref{fig:5} approach the same value (i.e., $\sim 0.15$) towards the highest RHIC and LHC energies is also supported by the limiting values ($\sim 1$) of the $K^-/K^+$ and $\pi^-/\pi^+$ ratios seen in Fig.~2 at the highest energies \cite{Andronic2009Horn}.

\begin{figure}[htbp]
    \centering
    \includegraphics[width=0.98\columnwidth]{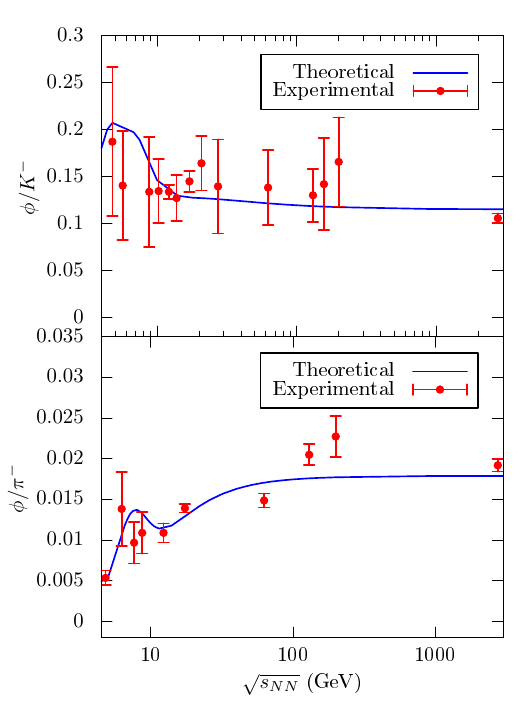} 
    \caption{Variation of $\Phi/\pi^-$, and $\Phi/K^{-}$ with $\sqrt{s_{NN}}$.}
    \label{fig:7}
\end{figure}

In Fig.~\ref{Fig:6},  we have shown the results of our calculations along with the experimental data for several other particle ratios, i.e., \( \Lambda (\bar{\Lambda})/\pi^- \), \( \Xi (\bar{\Xi})/\pi^- (\pi^+) \) and \( \Omega/\pi \) while in Fig.~\ref{fig:7} the variation of the $\Phi/\pi^-$ and $\Phi/K^{-}$ ratios with $\sqrt{s_{NN}}$ is shown.

We find that a reasonable fit is obtained in all these cases, thus highlighting the need for consideration of the strangeness imbalance effect in the system. The experimental data are taken from the references \cite{Adam2020,Abelev2010,Adamczyk2017B,Antii2003,RecentRF,Alt2008,Aggarwal2011S,Friese2004,E9172003,Nandi2020}.  The theoretical fits to the data sets give $\chi^2/\text{\textit{dof}}$ = 1.02 (1.89) for  \( \Lambda (\bar{\Lambda})/\pi^- \), 5.52 (5.27) for \( \Xi (\bar{\Xi})/\pi^- (\pi^+) \) and 1.12 for \( \Omega^-/\pi^- \). The larger values for \( \Xi (\bar{\Xi})/\pi^- (\pi^+) \) arise due to very small error bars associated with the LHC experimental data at 2.76 TeV. The corresponding values for the $\Phi/\pi^-$ and $\Phi/K^{-}$ are 2.61, 0.61  respectively.

\begin{figure}[H]
    \centering    \includegraphics[width=0.98\columnwidth]{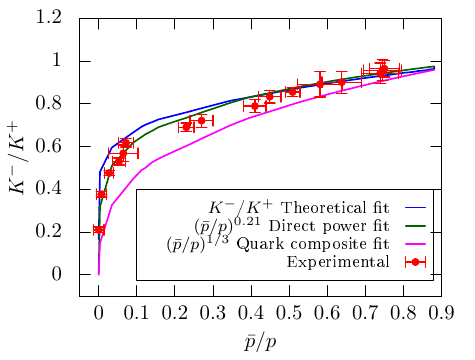} 
    \caption{Correlation between $K^-/K^+$ and $\bar{p}/p$ ratios}
    \label{fig:8}
\end{figure}

We have analyzed the correlation between $K^-/K^+$ and $\bar{p}/p$ ratios in URNNC over a wide energy range from 5 GeV to 2.76 TeV. The correlation is best described by a power-law type relationship as $K^-/K^+ = \left(\bar{p}/p\right)^\nu$. This can serve as validation for any thermal model as it indicates how kaon production relates to net baryon density.
In Fig.~\ref{fig:8}, we have shown the experimental data points along the with curves obtained from different approaches. The blue curve shows the result of our calculations. The pink curve is obtained by considering the light valence quark compositions, thus providing $\nu \approx 0.33$ but fits the data somewhat poorly. For comparison, if we represent the theoretical blue curve by a power-law fit, this provides $\nu \approx 0.17$, which contrasts sharply with the light valence quark composition approach. However, if we obtain a direct fit to the data using the power law, we get $\nu \approx 0.21$,  which is quite close to our theoretical results. This correlation is therefore well described by the VDW-HRG type approach in the framework of the grand canonical ensemble-based thermal model. Experimental data at different energies are taken from the references \cite{Bearden2003,Stiles2006,Das2014,Abelev2009}.
\vspace{-5mm} 
\vspace{-5mm}

\subsection{\texorpdfstring{ENERGY DEPENDENCE OF THERMAL FREEZE-OUT PARAMETERS ($T$, $\mu_B$, $\mu_s$, $\mu_Q$)}{ENERGY DEPENDENCE OF THERMAL FREEZE-OUT PARAMETERS (T, mu_B, mu_s, mu_Q)}}

\begin{figure}[H]
    \centering
    \includegraphics[width=0.98\columnwidth]{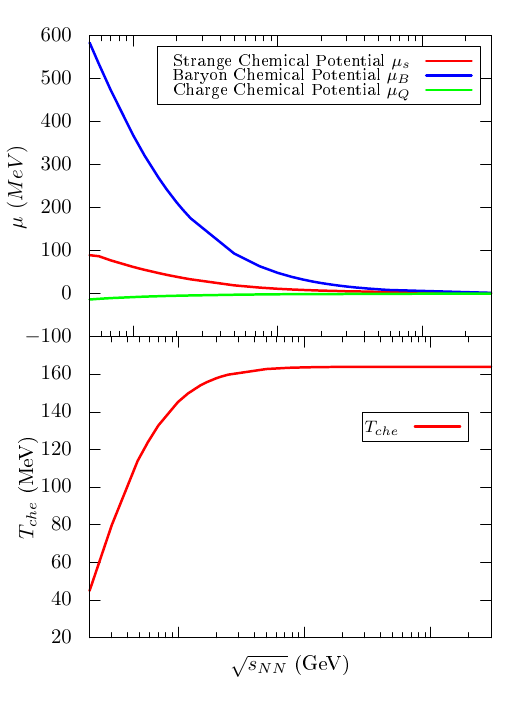} 
    \caption{Collision energy dependence of $\mu_B$, $\mu_s$, $\mu_Q$ and $T_{che}$.}
    \label{fig:9}
\end{figure}

In order to highlight the energy dependence of the freeze-out parameters of the model, we have shown in Fig.~\ref{fig:9}  the $\sqrt{s_{NN}}$ dependence of $T$, $\mu_B$, $\mu_s$, and $\mu_Q$ as extracted from the thermal fits of particle ratios. The energy dependence of $\xi_s$ has already been discussed in the previous section and shown in Fig.~\ref{fig:4}. The temperature curve saturates beyond $\sqrt{s_{NN}} \sim 50 \, \text{GeV}$ at 164 MeV. As discussed before this may indicate the onset of nuclear transparency around this energy, where there is lesser stopping of the colliding nuclear medium and the bulk production of secondary hadrons takes place during the evolution of the system between the receding nuclei \cite{Bashir2015}. The $\mu_B$ is seen to initially decrease rapidly with $\sqrt{s_{NN}}$ and later approach a negligibly small value beyond 200 GeV. The $\mu_s$, which is obtained by applying the strict strangeness conservation criteria in the hot and dense system formed in URNNC, shows a  decreasing trend from an initial value of $\sim$ 90 MeV to almost zero value beyond 200 GeV. Both $\mu_B$ and $\mu_s$ thus approach zero as the matter becomes more and more baryon symmetric with increasing $\sqrt{s_{NN}}$, which again is an indicator of the nuclear transparency effect setting in at the higher RHIC energies and further beyond up to LHC energies. In the same figure we have also shown the variation of the electric chemical potential $\mu_Q$. We find that there is an increasing trend from a very small initial value ($\sim$ -16 MeV) to almost zero towards higher energies as the system tends to become (almost) baryon symmetric.

\begin{figure}[htbp]
    \centering
    \includegraphics[width=0.98\columnwidth]{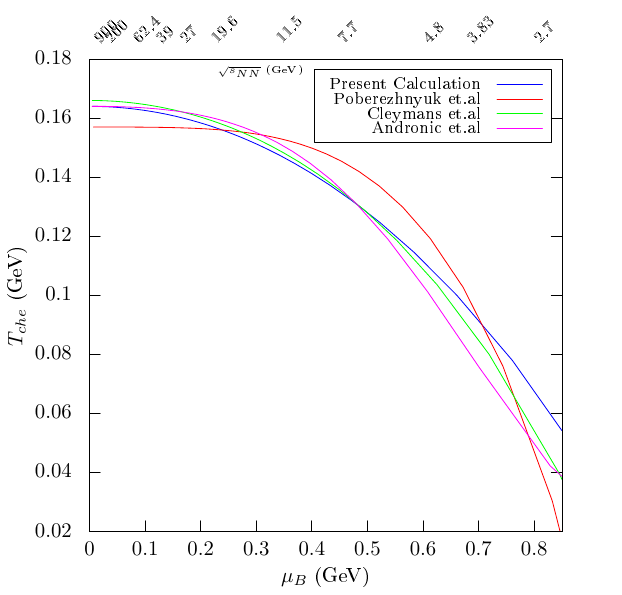} 
    \caption{Chemical freeze-out line from different calculations}
    \label{fig:10}
\end{figure}

Fig.~\ref{fig:10} illustrates a somewhat smooth parabolic dependence of the chemical freeze-out stages in the $T-\mu_B$ plane. The vertical axis represents the chemical freeze-out temperature, while the horizontal axis shows the corresponding baryon chemical potential. At RHIC energies, the chemical freeze-out temperatures $T_{\text{che}}$ ($\sim$ 160 - 164 MeV) are close to the hadron-QGP phase transition temperature predicted by lattice gauge theory \cite{Tiwari2013, Lysenko2024Chemical} thus indicating that freeze-out occurs almost immediately after the phase transition.
For the  sake of comparison, on the same graph are also shown some earlier results. The point-like case results of Cleymans et. al. \cite{Cleymans2006C} and Andronic et. al. \cite{Andronic2006} are shown along with the finite size case result of Poberezhnyuk et al. \cite{Poberezhnyuk2019Chemical} who have used the baryonic hard-core radius value as 0.59 fm. The results of Cleymans et. al. are in close agreement with our results while the results of Andronic et. al. and Poberezhnyuk et. al. show some deviations at small and intermediate collision energies. 
It may be noted that the extracted values of the freeze-out parameters in our case is dependent on the various particle ratios as obtained at different collision energies. On the other hand Poberezhnyuk et. al. have used the data on mean hadron multiplicities in various experiments to extract parameter values.

Our current model results nevertheless, in general, align somewhat closely with the findings of some other previous studies. The freeze-out temperatures ($T$) and baryon chemical potentials ($\mu_B$) for RHIC energies have been earlier reported, showing a range of $T$ values from 151 MeV to 167 MeV and $\mu_B$ values from 50 MeV to 20 MeV for higher RHIC energies, $\sqrt{s_{NN}} = 130 - 200 \, \text{GeV}$ \cite{Andronic2017Decoding,Cleymans2006C,Munzinger2001,Adams2005E,Baran2003,Florkowski2001fp,Chatterjee2015,Alba2020ChemicalFP,Lysenko2024Chemical,Poberezhnyuk2019Chemical, Cleymans2005C,Andronic2006,Florkowski2001,Tawfik2013, Tawfik2014}.

\section{SUMMARY AND CONCLUSION}
\label{Sec:sum}
In summary we have systematically applied the statistical VDW-HRG model in the framework of grand canonical ensemble to simultaneously describe several particle ratios in ultra-relativistic nucleus-nucleus collisions over a broad range of collision energy using a single set of model parameters. We have used a mean value of the parameter representing the attractive part of interaction obtained from the ground state properties of nuclear matter. The baryonic (antibaryonic) hard-core radius is treated as a free parameter. We have used an established ansatz form for the chemical freeze-out temperatures and baryon chemical potential. We have determined values of all freeze-out parameters from the (simultaneous) best theoretical fits obtained for the like-mass antibaryon to baryon ratio data sets using minimum chi squared fit method. We have explained the $K^-/K^+$, $\pi^-/\pi^+$ and proton to pion ratios quite well. Several other unlike-mass particle ratios (strange to non-strange) have also been well describes by the present calculation by introducing a strangeness imbalance factor ($\xi_s$). The experimentally observed correlation between $K^-/K^+$ and $\bar{p}/p$ ratios is well explained by our model calculations. 
A common feature of all ratios is that they tend to saturate at large collision energies. This is a clear indication of a nearly constant temperature and chemical potential (though vanishingly small) at higher collision energies, i.e., beyond 200 GeV. The decreasing chemical potential favours the production of antibaryons over baryons. We have illustrated the energy dependence of the freeze-out line and compared it with the previous results.
\vspace*{5mm} 

\section*{ACKNOWLEDGMENTS}
We would like to thank Jajati K. Nayak and Abdel Nasser Tawfik for their help in providing well-structured experimental data and their suggestions from time to time. The authors are also thankful to Jan-E Alam, Sandeep Chatterjee and Sabyasachi Ghosh for their valuable suggestions.

 \end{document}